\documentclass[10pt,a4paper]{article}

\usepackage{graphicx}
\usepackage{amsmath,amsfonts}%
\usepackage{tabularx}
\usepackage{enumitem}
\usepackage{xspace}
\usepackage{authblk}

\def\A{\mathcal{A}}
\def\P{\mathcal{P}}
\def\H{\mathcal{H}}

\def\Nset{\mathbb{N}}
\def\Esp{\operatorname{E}}
\def\Cov{\operatorname{Cov}}

\def\eg{{\it e.g.,\xspace}}
\def\ie{{\it i.e..,\xspace}}

\newcounter{maps}  
\newcounter{noise}

\title{On  the analysis  of signals  in a  permutation Lempel--Ziv  complexity -
  permutation Shannon entropy plane}

\author[1,*]{Diego M. Mateos* }
\author[2]{Steeve Zozor}
\author[3]{Felipe Olivarez}
\affil[1]{\normalsize Neuroscience and Mental Health Programme, Division of Neurology, Hospital for Sick Children. Canada. }
\affil[2]{\normalsize CNRS, GIPSA-Lab, Univ. Grenoble Alpes, France}
\affil[3]{\normalsize Instituto de Física, Pontificia Universidad Católica de Valparaíso, Chile.}
\affil[*]{Corresponding author: Diego M. Mateos, mateosdiego@gmail.com .}
\begin{document}

\maketitle

\begin{abstract}
  The  aim of  this paper  is to  introduce the  \textit{Lempel--Ziv permutation
    complexity} vs \textit{permutation entropy} plane  as a tool to analyze time
  series of  different nature.  This  two quantities make  use of the  Bandt and
  Pompe representation to quantify continuous-state time series. The strength of
  this plane is  to combine two different perspectives to  analyze a signal, one
  being statistic  (the permutation entropy)  and the other  being deterministic
  (the  Lempel--Ziv   complexity).   This  representation  is   applied  (i)  to
  characterize non-linear  chaotic maps, (ii) to  distinguish deterministic from
  stochastic  processes  and  (iii)  to  analyze  and  differentiate  fractional
  Brownian  motion from  fractional Gaussian  noise and  $K$-noise given  a same
  (averaged)  spectrum.   The results  allow  to  conclude  that this  plane  is
  ``robust'' to distinguish  chaotic signals from random signals,  as well as to
  discriminate between different Gaussian and nonGaussian noises.

\end{abstract}


\section{Introduction}
\label{Introduction:sec}

The signals coming  from the real world often have  very complex dynamics and/or
come from coupled  dynamics of many dimensional systems. We  can find an endless
number of examples arising from different fields. In physiology, one can mention
the reaction-diffusion  process in cardiac electrical  propagation that provides
electrocardiograms,  or the  dynamics underlying  epileptic electroencephalograms
(EEGs)~\cite{ZozBla03,  KanPut04}. The world  of finance  or social  systems are
also good examples of how ``complexity'' emerges from these systems~\cite{Art94,
  Tum84,  ShiFis84}.  One  important challenge  is then  to be  able  to extract
relevant information from these complex series \cite{Raj00, PonPro02, BroKin86}.

To address  this issue,  the researchers generally  analyze these  signals using
tools that come either from  probabilistic analysis, or from nonlinear dynamics.
The  idea in the  first approach  is to  measure the  spread of  the statistical
distribution underlying  the data,  to detect changes  in this  distribution, to
analyze the  spectral contents  of the  signals, etc. Among  the panel  of tools
generally employed,  that coming from  the information theory take  a particular
place~\cite{QuiArn00, RosBla01,  Sch00}. The second approach is  well suited for
signals having a deterministic origin,  generally non linear.  The tools usually
employed  come from  the  chaos world,  such  that the  fractal dimensions,  the
Lyapunov  exponents,  among  many  others~\cite{Raj00, WolSwi85},  or  from  the
concept  of  complexity  in  the   sense  of  Kolmogorov  (\eg  the  Lempel--Ziv
complexity)~\cite{Nag02, AboHor06, ZozRav05}

It  has  been  recently  proposed  to  analyze  time series  by  the  use  of  a
``complexity--entropy plane'',  exhibiting that the joint use  of two quantities
gives  a  richer information  about  the  series  than each  measure  separately
(see~\cite{VigBer03} for  the introduction of signal analysis  in an information
plane).  Such planes  have been  used in  the literature  in various  areas, for
instance to differentiate chaotic series from random signals~\cite{RosLar07}, to
characterize  chaotic  maps~\cite{RosOli13},  to  determine  the  authorship  of
literary    texts~\cite{RosCra09},     to    quantify    the     stock    market
inefficiency~\cite{ZunSor10},    or    for    the    analysis   of    the    EEG
dynamics~\cite{MonRos14}. In the mentioned  works, the complexity used are based
on  statistical measures,  for instance  the ``statistical  complexity measure''
making  use of the  entropy~\cite{LamMar04}, however  the  information obtained  from the
time series  is thus purely  statistical. In this  work, we propose to  use both
statistical and  deterministic measures namely,  the well known  Shannon entropy
and  Lempel--Ziv complexity~\cite{LemZiv76,  Sha48}. Additionally,  to calculate
the Lempel--Ziv  complexity as well as  the Shannon entropy, it  is necessary to
deal  with  time  series  taking  their  values  over  a  discrete  finite  size
alphabet~\footnote{ It is possible to calculate an entropy for continuous states
  signals, but  the estimation  of the so-called  differential entropy  from the
  data  is not an  easy task~\cite{BeiDud97,  LeoPro08}.}. Nevertheless,  in the
``real life'' we generally face to continuous-state data. Thus, the quantization
of the signal is a fundamental issue of the analysis.

Many  methods  to  quantify  continuous  data  exist  in  the  literature,  from
nonparametric estimators making  use of nearest neighbors or  graph lengths for
instance~\cite{BeiDud97, SchGra96,  LeoPro08, HerMa02, FrePom07},  to quantizers
based on  Parzen--Rosenblatt approaches (--for instance with  square kernel, \ie
using histograms)~\cite{Ros56,  Par62}. In this  paper, we focus on  an approach
proposed by Bandt and proposed a  decade ago, based on a multivariate trajectory
built  from  the  scalar  series,  \ie  an embedding,  and  the  so-called  {\em
  permutation  vectors} constructed  from this  embedding~\cite{BanPom02}.  More
precisely, for each point of the embedded trajectory, the vectors components are
sorted and  the values  are replaced by  their rank.  In their paper,  they used
these permutation vectors to propose what they called {\em permutation entropy},
that is nothing  more than the empirical estimate of the  Shannon entropy of the
permutation vectors~\cite{BanPom02}.  In the same vein, we proposed recently the
use  of  the Lempel--Ziv  complexity  applied  to  permutation vectors  for  the
analysis  of  continuous-states  sequences,   leading  to  what  we  named  {\em
  permutation  Lempel--Ziv complexity}~\cite{ZozMat14}.   The  same approach  is
envisaged here, but  instead of working with one of the  two single measures, we
propose to  study various  systems or sequences  in an  entropy-complexity plane
trying to characterize various type of behaviors from different time series.

The paper is organized as follows. Section~\ref{UncertaintyMeasures:sec} gives a
brief      introduction      on     the      measures      used     for      the
analysis. Section~\ref{Discretization:sec}  describes the permutation  method we
use  to  quantize  the series,  \ie  the  procedure  to obtain  the  permutation
vectors. Then, section~\ref{ChaosAndNoise:sec} presents characterizations of the
chaotic maps and  noises we analyze in the present work  while their analysis in
the  complexity--entropy plane is  given in  section~\ref{PLZC_PE:sec}. Finally,
some discussions are given in section~\ref{Discussion:sec}.


\section{Brief definitions of the uncertainty measures considered in the study}
\label{UncertaintyMeasures:sec}


\subsection{Shannon entropy}
\label{Shannon:subsec}

The concept of entropy was introduced in thermodynamics, quantum and statistical
physics by Boltzman, Gibbs or von Neumann among others~\cite{Bol64, Gib02, vNeu27,
  Nie52:v2,  Jay65,   MulMul09,  Pla15}  but   found  its  counterpart   in  the
communication   domain  through   the  seminal   work  of   Claude   Shannon  in
1948~\cite{Sha48}.  The  aim of Shannon was  to define a  measure of uncertainty
attached to a discrete-states random  variable under some axiomatics, namely (i)
the  invariance by  permutation  of  the probabilities  attached  to the  random
variable,  (ii) an  increase with  the  dimension of  the state  space when  the
distribution is  uniform, and (iii) a  recursivity property (ruling  the loss of
entropy  when joining two  states into  one).  These  so-called Shannon-Khinchin
axioms~\cite{Khi57} led to  the following definition of the  entropy $H[X]$ of a
discrete-states random variable, taking its outcomes in a discrete alphabet $\A$
of finite  size $\alpha =  |\A|$, with the  probability mass function  $p_X(x) =
\Pr[X = x], \, x \in \A$, as~\cite{Sha48, CovTho06}
\begin{equation}
H[X] = - \sum_{x \in \A} p_X(x) \log\left( p_X(x) \right).
\label{H_Shannon:eq}
\end{equation}    
In  the definition  of Shannon,  the logarithm  of  base 2  is used  and $H$  is
expressed in {\em bits}; the natural logarithm  can also be used and thus $H$ is
expressed in {\em nats}.

The so-called  Shannon entropy is  a functional of  the distribution of  $X$ and
does  not   depend  on  the  values   taken  by  random  variable   $X$.  It  is
straightforward to show that
\[
0 \le H[X]  \le \log |\A|,
\]
where $H[X] =  0$ is minimal when  $p(x) = \delta_{x,x_0}$ for a  given $x_0 \in
\A$ (all  the information about $X$ is  known) and $H[X] =  \log|\A|$ is maximal
when the distribution is uniform over $\A$. Thus, the logarithm $\log_\alpha$ of
base $\alpha$ should be preferred,
\[
h[X] = - \sum_{x \in \A} p(x) \log_\alpha(p(x)).
\]
so that the entropy is normalized. When  the dimension $d$ of $X$ is higher than
one,  one   sometimes  deals   with  the  entropy   per  number   of  components
$\frac{H[X]}{d}$ and for  an infinite sequence of variable  (or vector) with the
so-called entropy  rate that is the limit  of the entropy per  variable when the
length goes to the infinity.

The  definition of  the entropy  extended naturally  to the  case where  $|\A| =
+\infty$ (the entropy being then  unbounded and the maximal entropy distribution
when    some   moments    are   constrained    being   no    more    a   uniform
law)~\cite{CovTho06}. The extension to continuous random variable is not natural
and  was proposed  by  analogy with  the  discrete definition  by replacing  the
probabilities with the  probability density function and the  discrete sum by an
integral~\cite{CovTho06}.   However, the  so-called differential  entropy looses
several nice  properties of the  entropy, such that its  states-independence for
instance~\cite{CovTho06}.

In the context of signal analysis, when the probability distribution of the data
is unknown, the entropy cannot be evaluated and must be estimated from the data.
In the  discrete-state context, starting  from an observed sequence  of symbols,
the distribution can be estimated  through the frequencies of apparition of each
symbol.   Conversely,  in  the  continuous  state  context,  estimation  of  the
differential  entropy appears  to  be a  quite complicated  problem~\cite{Dob58,
  Par62, Vas76, KozLeo87, BeiDud97, LeoPro08, FrePom07}.


\subsection{Lempel--Ziv complexity}
\label{LempelZiv:subsec}

The  entropy  and  the  tools  associated with  (entropy  rate,\ldots)  allow  a
statistical characterization of a random variable and/or sequence. Conversely to
such an approach, Kolmogorov introduced  the notion of complexity of an observed
sequence, viewed  as a deterministic one (a  trajectory), to be the  size of the
minimal (deterministic) program (or algorithm) allowing to generate the observed
sequence~\cite[Chap.~14]{CovTho06}. This notion is closely linked to that of the
compressibility of a sequence. Later on,  Lempel and Ziv proposed to define such
a  complexity  for the  class  of  ``programs''  based on  recursive  copy-paste
operators~\cite{LemZiv76}. Their approach precisely  gave rise to the well known
algorithms of compression such  that the famous 'gzip'~\cite{CovTho06, ZivLem77,
  ZivLem78, WynZiv89}.

To  be more  precise,  let us  consider  a finite-size  sequence  $ S_{0:T-1}  =
S_0...S_{T-1}$  of size  $T$, of  symbols  $S_i$ that  take their  values in  an
alphabet  $\A$ of  finite size  $\alpha =  |\A|$.  The  definition of  the first
version  Lempel--Ziv  complexity \cite{LemZiv76}  lies  in  the two  fundamental
concepts of reproduction and production:

\begin{itemize}
\item {\em Reproduction}:  it consists of extending a  sequence $S_{0:T-1}$ by a
  sequence  $Q_{0:N-1}$  via recursive  copy-paste  operations,  which leads  to
  $S_{0:_T+N-1} = S_{0:T-1} Q_{0:N-1}$, \ie where the first letter $Q_0$ is in
  $S_{0:T-1}$, let us  say $Q_0 = S_i$,  the second one is the  following one in
  the extended sequence of size $T+1$, \ie $Q_1 = S_{i+1}$ , etc.: $Q_{0:N-1}$
  is a subsequence  of $S_{0:T+N-2}$. In a sense, all  of the ``information'' of
  the extended sequence $S_{0:T+N-1}$ is in $S_{0:T-1}$.
\item {\em  Production}: the extended sequence  $S_{0:T +N-1}$ is  now such that
  $S_{0:T+N-2}$ can  be reproduced  by $S_{0:T-1}$, but  the last symbol  of the
  extension can either  follow the recursive copy-paste operation  (thus we face
  to a  reproduction) or  can be ``new''.   Note thus  that a reproduction  is a
  production,  but  the  converse is  false.   Let  us  denote a  production  by
  $S_{0:T-1} \Rightarrow S_{0:N+T-1}$.
\end{itemize}

Any sequence can  be viewed as constructed through  a succession of productions,
called  an  history  $\H$.  For  instance,  an  history  of $S_{0:T-1}$  can  be
$\H(S_{0:T-1}): \emptyset \Rightarrow  S_0 \Rightarrow S_{0:1}\Rightarrow \cdots
\Rightarrow  S_{0:T-1}$.  The  number the  productions used  for  the generation
$C_{\H(S_{0:T-1})}$ is here equal to the size of the sequence. A given sequence
does not have  a unique history and in the spirit  of the Kolmogorov complexity,
Lempel and Ziv were interested by  the optimal history, \ie the minimal number
of production  necessary to  generate the sequence.   The size of  the shortest
history  is the  so-called Lempel--Ziv  complexity, denoted  as  $C[S_{0:T-1}] =
\min_{\H(S_{0:T-1})}    C_{\H(S_{0:T-1})}$~\cite{LemZiv76}.     In   a    sense,
$C[S_{0:T-1}]$  describes the  ``minimal''  information needed  to generate  the
sequence $S_{0:T-1}$ by recursive copy-paste operations.

Clearly, $2  \le C[S_{0:T-1}] \le  T$.  But a  tricky analysis of  the possibles
sequences of length $T$ of symbols  of finite size alphabet allows to refine the
upperbound~\cite{LemZiv76}, so that, for any sequence,
\[
2 \:  \le \: C[S_{0:T-1}]  \le \frac{T}{(1-\varepsilon_T) \log_\alpha  T} \qquad
\mbox{where}  \qquad \varepsilon_T  = \frac{2  \, (1  +  \log_\alpha \log_\alpha
  (\alpha T))}{\log_\alpha T}
\]
Thus, sometimes, the ``normalized'' Lempel--Ziv complexity
\[
c[S_{0:T-1}] = \frac{C[S_{0:T-1}] \, \log_\alpha T}{T}
\]
is considered. Asymptotically  with the length of the sequence,  $c$ goes to the
interval $[0 \, , \, 1]$.

The Lempel--Ziv complexity has various  properties. Among them, it is remarkably
connected  to the  Shannon entropy  rate when  dealing with  sequences randomly
drawn: from  an ergodic sequence\footnote{A  sequence $S_{0:T-1}$ is  ergodic if,
  for any function  $g$, $\sum_{k \ge 0} g(S_k)$ converges  (almost surely) to a
  deterministic value.},
\begin{equation}
\lim_{T \to \infty} c[S_{0:T-1}] \: = \: \lim_{T \to \infty} \frac{h[S_{0:T-1}]}{T}
\label{LZC_Shannon:eq}
\end{equation}
with  probability 1 (see~\cite{LemZiv76,  Han89}). Note  that in  the stationary
case (the statistics are invariant by time translation), the joint entropy being
lower  or equals  to the  sum of  the individual  entropies~\cite{CovTho06}, the
limit is upperbounded by the entropy of the symbols in the sequence.


\section{From continuous-state signals to discrete states representations}
\label{Discretization:sec}

As previously introduced, the aim of the present paper is to analyze time series
in  a complexity-entropy plane,  namely the  Lempel--Ziv complexity  and Shannon
entropy plane.  However, as just seen, the Lempel--Ziv complexity can be defined
only  for discrete-states sequences.   Concerning the  Shannon entropy,  we have
also  seen  that   even  if  definitions  exist  for   both  discrete-state  and
continuous-states  random  variables, it  is  more  adapted  to the  uncertainty
description  of  discrete-state variables  (definition  based  on an  axiomatic,
estimations problems in the continuous-state context).

By the  way, when dealing with  continuous-states data, that is  the most natural
case in  various contexts, before any  analysis, an observed sequence  has to be
quantized.   Various  methods can  be  envisaged,  both  having impacts  on  the
interpretation of  the uncertainty  measure (entropy, complexity)  associated to
the  hence quantized  sequence (see  \eg~\cite{KasSch87, ZozMat14}).  We focus
here on an  approach based on the so-called permutation vector,  that was at the
heart   of   the   so-called   permutation   entropy   proposed   by   Bandt   and
Pompe~\cite{BanPom02}.


\subsection{Permutation vectors}
\label{PermutationVectors:subsec}

The Bandt  and Pompe  approach~\cite{BanPom02} seems to  take its origin  in the
study of chaos, and more specifically through the famous Takens' delay embedding
theorem~\cite{Tak81}.  The  core of this theorem concerns  the reconstruction of
the state  trajectory of a dynamical system  from the observation of  one of its
states.  To fix the ideas, consider a real-valued discrete-time series \ $\{ X_t
\}_{t \in  \Nset}$ \ and two integers  \ $d \geq 2$  \ and \ $\tau  \geq 1$, and
from the series,  let us then define a trajectory  in the $d$-dimensional space,
also called {\em embedding}, $\{ Y_t^{d,\tau} \}_{t \geq (d-1) \tau}$, \ as:
\begin{equation}
Y^{d,\tau}_t =[X_{t-(d-1)\tau} \quad \cdots \quad X_{t-\tau} \quad X_t]
\end{equation}  
where dimension  $d$ is  called {\em embedding  dimension}, and where  $\tau$ is
called {\em delay}.  In the domain  of chaos analysis, the Takens' theorem gives
conditions  on  $d$  and  $\tau$ such  that  the  embedding  $Y^{d,\tau}_t$
preserves  the  dynamical properties  of  the  full  dynamic of  the  underlying
system. This  point goes beyond the  scope of the  present paper, thus, we  do not
enter into details and let the readers refer to Ref.~\cite{Tak81, Rob11}.

Now,  the idea  of Bandt  and  Pompe to  map continuous-state  time series  into
discrete-states   one   consists  in   replacing   each   component  of   vector
$Y^{d,\tau}_t$  by its rank  when the  components are  sorted (\eg  in ascending
order). Such a  discrete-state vector, taking its values  over the alphabet $\P$
issued from the  ensemble of the permutation of  $\{0,\ldots,d-1\}$, of cardinal
$|\P|   =   d!$   is   called   {\em  permutation   vector}   and   is   denoted
$\Pi\left(Y^{d,\tau}_t\right)$ in the  following. As an illustration, $\Pi\left(
  [.55  \quad 1.7 \quad  -.45] \right)  = [1  \quad 2  \quad 0]$.   As mentioned
in~\cite{ZozMat14},   such   a  quantization   is   somewhat   similar  to   the
$\Sigma\Delta$  quantization  that  consists   in  quantizing  in  one  bit  the
difference between  the symbols $X_t$ and  a prediction of this  symbol from the
past~\cite{GerGra92}.   This is  roughly  similar  to quantize  in  one bit  the
variations of signal: 0 for decreasing  steps and 1 for increasing steps.  It is
precisely  the case  dealing  with $\Pi\left(Y^{2,1}_t\right)$,  leading to  $[0
\quad 1]$  for increasing  steps, and  $[1 \quad 0]$  for decreasing  steps.  As
mentioned   in  several   papers~\cite{AmiKoc06,   Ami10,  AmiZam07,   RosCar12,
  RosOli13},  the frequencies  or organization  of  patterns in  a sequence  can
reveal a chaotic behavior versus a random one.


\subsection{Permutation uncertainty and complexity measures}
\label{PermutationEntropyComplexity:subsec}

In  their paper~\cite{BanPom02}, Bandt  and Pompe  defined the  {\em permutation
  entropy} as  the Shannon entropy  of the empirical distribution  associated to
the permutation vectors, \ie where the probabilities are the frequencies of each
possible  permutation vector  in  the sequence  $\left\{ \Pi\left(  Y^{d,\tau}_t
  \right)  \right\}_t$.  When dealing  with  time-series  issued  from a  random
process, this is  nothing more than an estimation of the  Shannon entropy of the
permutation  vectors process.  In the  following,  we use  the terminology  {\em
  permutation Shannon entropy} to be  more precise. Moreover, to distinguish the
permutation entropy  as defined by  Bandt and Pompe  to the formal entropy  of a
random permutation vector, we will use the notation
\[
\widehat{h}^{d,\tau}[X_{0:T-1}]   =  -   \sum_{\pi  \in   \P}  \widehat{f}_X^{\,
  d,\tau}(\pi) \log_{d!} \widehat{f}_X^{\, d,\tau}(\pi)
\]
where  $\widehat{f}_X^{\, d,\tau}(\pi)$  is  the proportion  (frequency) of  the
permutation vector $\pi$ in the sequence $\left\{ \Pi\left( Y^{d,\tau}_t \right)
\right\}_t$.

\

Similarly   to   the   approach   of    Bandt   and   Pompe,   in   a   previous
work~\cite{ZozMat14}, we proposed the  Lempel--Ziv complexity of the permutation
vector sequence,  as a tool  for the analysis  of complex time series.  We named
this tool {\em permutation Lempel--Ziv complexity}, which is denoted
\[
\widehat{c}^{\,d,\tau}[X_{0:T-1}] =  c\left[\, \Pi \left(  Y^{d,\tau}_{(d-1) \tau}
  \right) \, \cdots \, \Pi \left( Y^{d,\tau}_{T-1} \right) \, \right]
\]
in the sequel.

\

In the  present paper, we propose to  analysis time series in  the hence defined
{\em  permutation  Lempel--Ziv complexity  --  permutation  Shannon entropy},  by
representing    a    sequence     $X_{0:T-1}$    by    a    ``point''    $\left(
  \widehat{h}^{d,\tau}[X_{0:T-1}]  \:   ,  \:  \widehat{c}^{\,d,\tau}[X_{0:T-1}]
\right)$ in a 2D-plane. The motivations lie on the following observations:
\begin{itemize}
\item  As noted by  several authors~\cite{AmiKoc06,  Ami10}, in  various chaotic
  context, the presence of forbidden patterns can reveal the chaotic behavior of
  a  sequence  since,  in general,  there  is  no  forbidden pattern  in  random
  sequences. As an example, for the  logistic map $X_{t+1} = 4 X_t (1-X_t)$, for
  $d  =  3$  and  $\tau  =  1$,   the  pattern  $[2  \quad  1  \quad  0]$  never
  appears. However, this is not always  the case: there are chaotic maps without
  forbidden    patterns    and    conversely   noises    exhibiting    forbidden
  patterns~\cite{AmiZam07, RosCar12}.   Thus, the chaotic aspect  can be revealed
  by  the  time  organization  of   the  permutation  vectors  rather  by  their
  frequencies  of  occurrence, \ie  by  the  permutation Lempel--Ziv  complexity
  rather than by the permutation Shannon entropy.
\item  From the  relations Eqs.~\eqref{LZC_Shannon:eq}  for  (sufficiently long)
  stationary and ergodic random sequences, the normalized complexity reaches the
  entropy rate of  the sequence. From the fact that  for random variables $X_i$,
  $H(X_1,\ldots,X_n)  \le \sum_i  H(X_i)$~\cite{CovTho06}, the  entropy  rate is
  always  less than individual  entropies. In  other word,  one may  expect that
  $\widehat{c}^{d,\tau}  \le \widehat{h}^{d,\tau}$  with equality  for sequences
   of permutation vectors with independent and identically distributed samples.
 \item Moreover, if  in a parametrized family of noise  series, the entropy rate
   (of the  permutation vectors) is linked  to the individual  entropy through a
   function   of    the   parameter,   the    curve   $(\widehat{h}^{d,\tau}   ,
   \widehat{c}^{d,\tau})$ is expected to be close to this curve
 \item From the previous remark, one may expect that in chaotic sequences, for a
   given  permutation entropy,  the  complexity  will be  lower  than for  noisy
   sequences  due  to the  temporal  organization  governed  by a  deterministic
   dynamics.
\end{itemize}

In other words,  it is expected that various  kind of noise and of  chaos can be
finer characterized (separated) in such  a plane, by distinguished in some sense
the  part of  algorithmic complexity  and  the part  of statistical  uncertainty
contained in a time series.

Finally, note that  the proposed analysis of a series  applies also dealing with
intrinsic  vector  series.   In  this  case,  as  done  in~\cite{ZozMat14},  the
permutation vectors are issued from the  vector of the trajectories (and thus no
embedding is done previously to permutation procedure).

\

Let us now turn  to the various type of time series we  aim at studying with the
proposed approach.


\section{Characterization of chaotic maps and noises}
\label{ChaosAndNoise:sec}


\subsection{Chaotic maps}
\label{ChaoticMaps:subsec}

In the  present work,  we consider 26  chaotic maps  described by Sprott  in the
appendix of his book~\cite{Spr03}.  These chaotic maps are grouped as follows.
\begin{itemize}
\item   \textit{Conservative  Maps:}   In  contrast   to   dissipative  systems,
  conservative systems have some conserved quantities, such as mechanical energy
  or angular momentum. In this  case, the phase-space volume is preserved. These
  systems arise  naturally in the \textit{Hamiltonian}  formulation of classical
  (Newtonian) mechanic,  and they are also  called \textit{Hamiltonian systems}.
  In this paper, we analyze the  following conservative maps (we will label each
  case by its number in the plane of analysis in next section):
  \begin{enumerate}[label=(\arabic*)]
  \item The Arnold's cat map
  \item The chaotic web map
  \item The Chirikov standard
  \item The Gingerbreadman
  \item The H\'enon area-preserving quadratic
  \item The Lorenz three-dimensional chaotic map
  \setcounter{maps}{\value{enumi}}
  \end{enumerate}
\item \textit{Dissipative  maps:} Dissipative mechanical systems  are systems in
  which  mechanical energy is  converted (or  \textit{dissipated}) into  heat. A
  consequence is that the phase-space volume contracts~\cite{Spr03}.
  In  this paper,  we  analyzed  the following  dissipative  maps:
  \begin{enumerate}[label=(\arabic*)]
  \setcounter{enumi}{\value{maps}}
  \item The H\'enon map
  \item The Lonzi map
  \item The Delayed logistic map
  \item The Tinkerbell map
  \item The Holmes cubic map
  \item The dissipative standard map
  \item The Ikeda map
  \item The Sinai map
  \item The discrete predator prey map
  \setcounter{maps}{\value{enumi}}
  \end{enumerate}
\item \textit{Non-inverted maps:} An iterated map is called noninverted, when in
  a  sequence,  each iterate  $X_n$  has two  preimages  $X_{n-1}$  that do  not
  coincide.  Consequently, one  bit of information (a factor of  2) is lost with
  each iteration  since there is no way  to know from which  preimage each value
  came. This exponential loss of information is equivalent to exponential growth
  of   the  error   in  the   initial  condition   that  is   the   hallmark  of
  chaos. Noninvertibility is necessary for chaos in one-dimensional maps but not
  for maps  in higher dimension~\cite{Spr03}.   Here, we analyzed  the following
  non inverted maps:
  \begin{enumerate}[label=(\arabic*)]
  \setcounter{enumi}{\value{maps}}
  \item The lineal congruential generator
  \item The cubic map
  \item The Cusp map
  \item The Gauss map
  \item The logistic map
  \item The Pinchers map
  \item The Ricker's population model
  \item The sine circle map
  \item The sine map
  \item The Spence map
  \item The tent map
  \setcounter{maps}{\value{enumi}}
  \setcounter{noise}{\value{enumi}}
  \end{enumerate}
\end{itemize}
For all the  maps presented above, we use the  parameters and initial conditions
expressed  in~\cite{Spr03}.  For  more  detail about  each map  see~\cite{Spr03,
  RosLar07, RosOli13}


\subsection {Random sequences}
\label{Noise:subsec}

As well  known, a real valued random  signal $X(t)$ is characterized  by all the
joint distribution of  $(X(t_1),\ldots,X(t_n))$ for any $n \in  \Nset^*$ and any
set  of times  $(t_1,\ldots,t_n)$.  When the  signal  is Gaussian,  that is  any
$(X(t_1),\ldots,X(t_n))$  has a  Gaussian distribution,  the signal  is entirely
described by its  first two order moments, namely its  mean $\Esp[X(t)]$ and its
covariance  function $C_X(t,s)  = \Cov(X(t),X(s))$.  Let us  also recall  that a
signal is said stationary if the  statistics are invariant by any time shift. In
particular, the covariance  $C_X(t,s)$ depends only on $u  = s-t$, $C_X(t,t+u) =
C_X(u)$.  In  general, such  a signal  is characterized more  likely by  it power
spectral density, given  by the Fourier transform $\Gamma(f)$  of the covariance
(Wiener-Kintchin theorem).

In  the sequel,  we  aim at  studying  the following  Gaussian and  nonGaussian,
stationary of nonstationary noises:
\begin{itemize}
 
\item $K$-noises, where  the power spectrum takes the  form $1/f^k$. Noises with
  such   a  power-law   spectrum   are   widely  found   in   nature,  like   in
  physics~\cite{DutHor81},          in          biology~\cite{Hos81},         in
  astrophysics~\cite{WesShl90}  among  other  domains.   Such  a  noise  is  not
  necessarily  Gaussian.   In particular,  in  this  paper,  we focus  on  noise
  generated  through the algorithm  described in~\cite{RosLar07}  that basically
  consists in  (i) generating  a pseudo random  sequence of  independent samples
  with uniform  probability distribution  and zero mean  value, (ii)  taking the
  Fourier transform, (iii) multiplying  this Fourier transform by $f^{-k/2}$ and
  symmetrizing the  result so as to  obtain a real function  (iv) performing the
  inverse  Fourier  transform  and  discarding the  small  imaginary  components
  produced  by numerical  approximations. the  obtained sequence  appears  to be
  nonGaussian~\cite{RosLar07}.
  \addtocounter{noise}{15}
  \begin{enumerate}[label=(\arabic{enumi})-(\arabic{noise}),leftmargin=1.5cm]
  \setcounter{enumi}{\value{maps}}
  \item We concentrate here on $k = 0.25 \times n, \: n = 0, \ldots 14$.
  \end{enumerate}
\item  Standard fractional  Brownian  motion (FBM).   Such  a Gaussian  process
  $B_H(t)$ is  non-stationary and parametrized  by a quantity  $H \in (0  , 1)$,
  called  Hurst  exponent, and  has  the  covariance  function $$C_{B_H}(t,s)  =
  \frac{1}{2} \Big( |t|^{2H}  + |s|^{2H} - |t-s|^{2H} \Big).$$  This process was
  introduced  by  Kolmogorov~\cite{Kol40}   and  studied  by  the  climatologist
  Hurst~\cite{Hur51} or  later on by Mandelbrot and  Van Ness in~\cite{ManNes68}
  for  modeling fractals  for instance.   In this  last reference,  the authors
  defined such  a process through  a Weyl integral,  with in general  the choice
  $B_H(0) = 0$ almost surely (see~\cite{ManNes68, Fla89} for more details).  The
  FBM increments $B_H(t)-B_H(s)$ are stationary and the process is self-similar,
  \ie $B_H(at)$  has the  same distribution than  $|a|^H B_H(t)$  \cite[and ref.
  therein]{ZunPer07, Fla89, Ber94, SamTaq94,  Fed88}.  These processes exhibit a
  very  rich behavior  depending on  $H$: For  $H =  \frac12$, one  recovers the
  standard Brownian  motion (limit process of  the random walk); For  $H > 1/2$,
  the process exhibits  persistency in the sense that a  given trend or increment
  sign  in the  past  tends to  persist in  the  future (the  increments have  a
  positive  correlation)  and  the  process  exhibits  long  range  dependence;
  Conversely, for $H < 1/2$, the process is antipersistent in the sense that the
  trends from past to future are more likely to be opposite (the increments have
  a negative  correlation).  Finally, note that  the spectrum\footnote{Since the
    process is, dealing with spectrum has  no sense in itself.  However, one can
    consider    it   through    the   Wigner-Ville    spectrum,    averaged   in
    time~\cite{Fla89}, which would be the  spectrum estimated from a sample path
    for      instance.}       of     a      FBM      is     proportional      to
  $\frac{1}{f^{2H+1}}$~\cite{Fla89, MolLiu97}.
\item  Fractional  Gaussian noise  (FGN).   Such a  process  is  defined as  the
  increments of a  FBM~\cite{Sam07}, as $$G_H(t) = B_H(t+1)-B_H(t)$$  Due to the
  stationarity  of  increments   of  FBM,  a  FGN  is   stationary  and  it  is
  straightforward to show that its covariance function is $$C_{G_H}(u) = \frac12
  \Big( |u+1|^{2H}  - 2  |u|^{2 H}  + |u-1|^{2 H}  \Big).$$ Note  that for  $H =
  \frac12$  the  correlation  function  vanished  for non-zero  lags  $u$.   Thus
  $G_{\frac12}$  corresponds to Gaussian  white noise.   Finally, note  that the
  spectrum of FGN is proportional to $\frac{1}{f^{2H-1}}$~\cite{MolLiu97}.
\end{itemize}
To generate time series from FBM and FGN, we used the algorithm proposed by Abry
and  Sellan~\cite{DavHar87, AbrFel96}.   By nature,  the sequences  generated by
this algorithm are discrete-time  approximation of the continuous-time sequence,
what  is precisely needed  to be  able to  analyze such  sequences in  the plane
previously introduced.


\section{Permutation Lempel--Ziv Complexity vs Permutation Entropy plane}
\label{PLZC_PE:sec}

To illustrate how the  permutation Lempel--Ziv complexity vs permutation Shannon
entropy plane  can reveal characteristics of  a time series,  we analyze through
this    plane     the    chaotic     and    random    series     described    in
section~\ref{ChaosAndNoise:sec}.  Our purpose  is to  exhibit that  the proposed
plane  allows to  distinguish  the random  from  chaotic signals,  but also  to
separate Gaussian  and nonGaussian processes  with the same spectrum  over their
``degree of correlation''.

\subsection{Chaos and $K$-noises analysis}
\label{Chaos_Knoise_Analysis:subsec}

\begin{figure}[htbp]
\centerline{\includegraphics[scale=0.6]{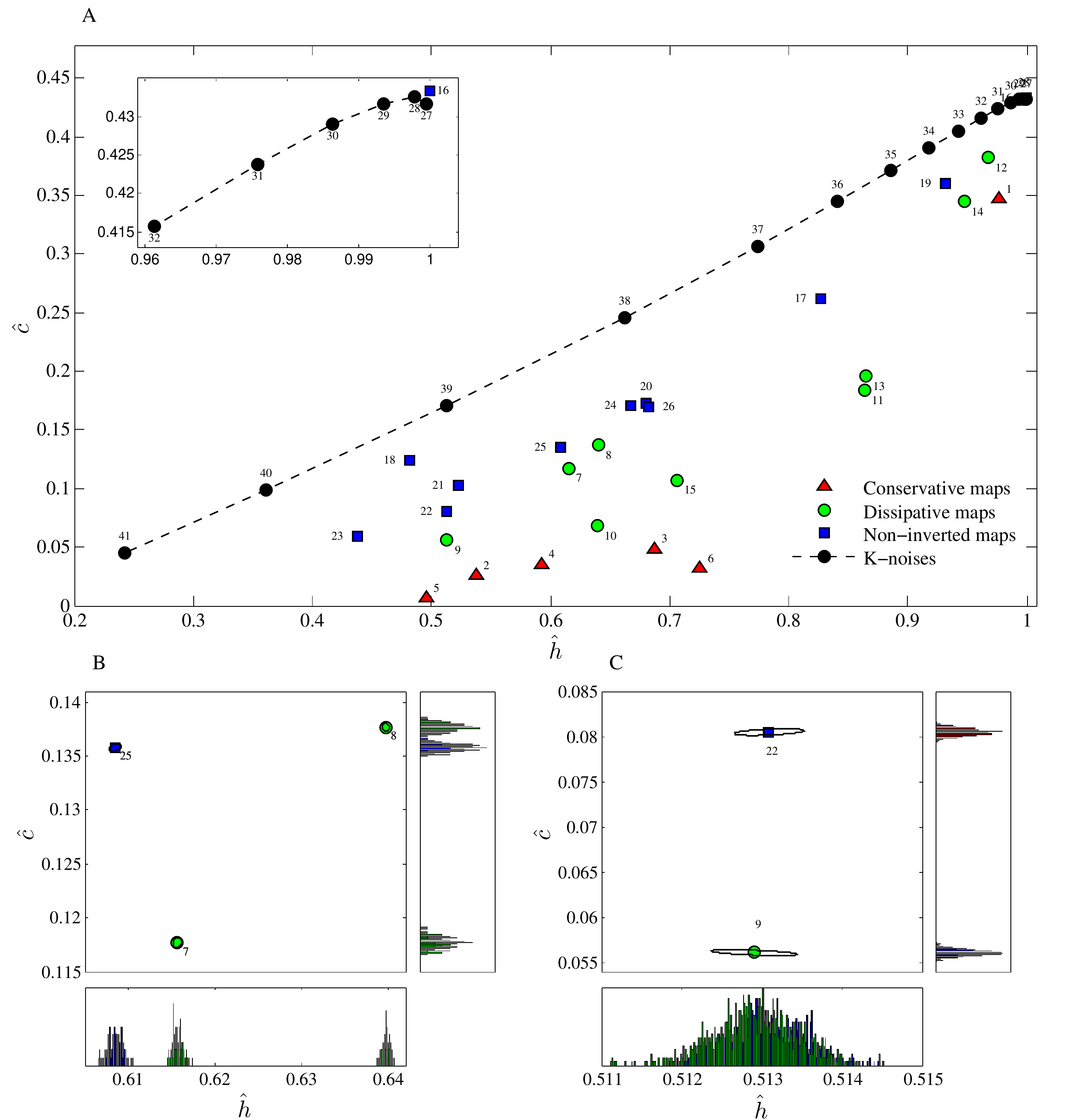}}
\caption{  A)   Localization  in  the  permutation   Lempel--Ziv  complexity  vs
  permutation Shannon entropy  plane of the chaotic series  and $K$-noise series
  considered in  the present  work, for the  parameters $d=5$ and  $\tau=1$. For
  each  case, we generated  $N =  4.10^4$ times  series of  $L =  10^6$ samples,
  initializing  each  of them  randomly.   The  labels  corresponds to  that  of
  section~\ref{ChaosAndNoise:sec}.  All  chaotic maps are  clearly separate from
  the $K$-noises. B-C)  Zoom in a zone containing  the coordinates for specifics
  chaotic  maps.   The  inferior  and  lateral plots  depict  the  corresponding
  histograms  of the  values taken  by each  measure separately  using the  $N =
  4.10^6$ snapshots.   The ellipses corresponding to the  standard deviation are
  small. Similar result are obtained for the parameters $d=4,6$ and $\tau=1$ }
\label{mapa_CyR:fig}
\end{figure}

For each of the 41 times series labeled in the previous section ((1) to (26) for
the chaotic maps, and (27) to (41)  for the $K$-noise) we generated $N = 4.10^4$
times series  of $L =  10^6$ samples, initializing  each of them  randomly.  For
each  snapshot,  of each  series,  we  computed  the corresponding  sequence  of
permutation  vectors   (choosing  the  parameters  $(d,\tau)$),   and  thus  the
permutation Shannon  entropy $\hat{h}^{d,\tau}$ and  the permutation Lempel--Ziv
complexity   $\widehat{c}^{d,\tau}$.   Figure~\ref{mapa_CyR:fig}A   depicts  the
points   $\left(   \langle   \hat{c}^{d,\tau}   \rangle_N  \,   ,   \,   \langle
  \hat{h}^{d,\tau}  \rangle_N \right)$ where  $\langle \cdot  \rangle_N$ denotes
the  averaged  quantities  over  the  $N$  realizations  and  where  the  chosen
parameters are $(d,\tau) = (5,1)$.  We also tested the embedding dimensions $d =
4$  and  $d =  6$  (with the  same  lag  $\tau =  1$);  the  repartition of  the
coordinates  in   the  complexity-entropy  plane   is  similar,  and   thus  the
observations are  robust to dimension $d$. Figure~\ref{mapa_CyR:fig}B  and C are
zooms in a zone containing the coordinates for specifics chaotic maps.  The dots
represent  the  mean values,  as  for  figure~\ref{mapa_CyR:fig}A. The  ellipses
represent  the dispersions  of  the values  over  the snapshots  via the  sample
covariance  matrix  $\widehat{C}_{\widehat{h},\widehat{c}}$  computed  from  the
data,  \ie  an ellipse  corresponds  to  $\left[  \widehat{h} \quad  \widehat{c}
\right]  \, \widehat{C}_{\widehat{h},\widehat{c}}  \,  \left[ \widehat{h}  \quad
  \widehat{c} \right]^t  = 1$.  The inferior and  lateral histograms  depict the
corresponding histograms of the values taken by each measure separately using $N
= 4.10^4$ snapshots.  We can see the benefits of studying the series on a plane.

One can observe in these figures that the classes of series are unseparable when
dealing with  a single  measure, while using  a statistical and  a deterministic
measure simultaneously, one can better distinguished their respective nature.

First of all, the  complexity-entropy coordinates corresponding the sequences of
noise are  remarkably aligned on  a line, while  that of the  chaotic sequences
separates  clearly from  this line.  As  known, the  entropy rate  of a  process
decreases with  the correlation of the  process. The observed  alignment of the
points issued from  the sequences of the $K$-noise reveals  that such a behavior
remains for the permutation vectors process, and moreover that the dependence is
more or  less linear vs $k$,  so that, due  to the asymptotic behavior  of these
stationary ergodic sequences, the permutation complexity as well.  This behavior
is singularly different for the chaotic sequences.
It is  important to note  however a  small deviation from  the line and  a small
decrease   of  the  complexity   for  very   small  $k$   (see  the   insert  of
Fig.~\ref{mapa_CyR:fig}A). This  effect can be explained by  the quantization of
the data through  the permutation vectors. Indeed, we observe  that when $d \geq
10$, this deviation from the  line tends to disappear\footnote{Note however that
  for high  dimension, the size of the  alphabet $|\P| = d!$  becomes very large
  and thus the sequences to be analyzed must be drastically large to insure that
  both the permutation complexity and permutation entropy have a meaning}.

Regarding  the chaotic  maps, as  already observed,  the  representative average
coordinates clearly  separate from the ``noise-line'', and  always is positioned
below this  line. The separation  from the line  is indeed a consequence  of the
deterministic  dynamic  underlying  such   processes,  mechanisms  that  are  of
relatively low complexity. Thus, for  the same single entropy, chaotic sequences
have  a lower  complexity than  noise. A  notable exception  lies in  the lineal
congruential  map  (16) (see  the  insert  of figure~\ref{mapa_CyR:fig}A).  This
exception  can  be explained  by  the  pathological  characteristic of  this  map.
Indeed, sequences generated by this map are often used to generate pseudo-random
sequences  and  share   a  huge  number  of  characteristic   of  purely  random
sequences~\cite[Chap.~5]{KinRee98}.  Moreover,  analyzing the correlation  (in a
deterministic sense),  it appears  that the correlation  of the samples  is very
small, explaining why the coordinate  entropy-complexity of this map is so close
to that of the Gaussian white noise.

Finally,  note  that   the  chaotic  maps  are  relatively   well  separated  in
``clusters''  regarding their  classification ``non  inverted'', ``dissipative''
and ``conservative''. This observation suggests  that the analysis of a sequence
in a permutation Shannon entropy and permutation Lempel--Ziv complexity plane is
powerful to finely characterize the class of such sequences. To further analyze
the results,  let us  mention that  the proposed map  allows to  distinguish the
H\'enon area-preserving quadratic map (5) from the delayed logistic map (9), the
Pinchers map  (21) from the Gingerbreadman map  (4) or the Spence  map (25) from
the  H\'enon  map  (7), maps  that  are  less  distinguishable using  the  plane
previously  proposed in  the literature~\cite{RosOli13}.  The same  situation is
observed between chaotic maps and $K$-noises as for example dissipative standard
map (12) and the correlated noise with ($K=1.25$). These differences are measure
by the  implementation of a non  statistical measures such  that the Lempel--Ziv
complexity, demonstrating  that the plane  of analyze we  propose here can  be a
good  alternative when  sequences  are  not separable  in  the plane  previously
proposed in the literature (and conversely).

\subsection{$K$-noises, FBM and FGN analysis }

\begin{figure}[htbp]
\centerline{\includegraphics[scale=0.45]{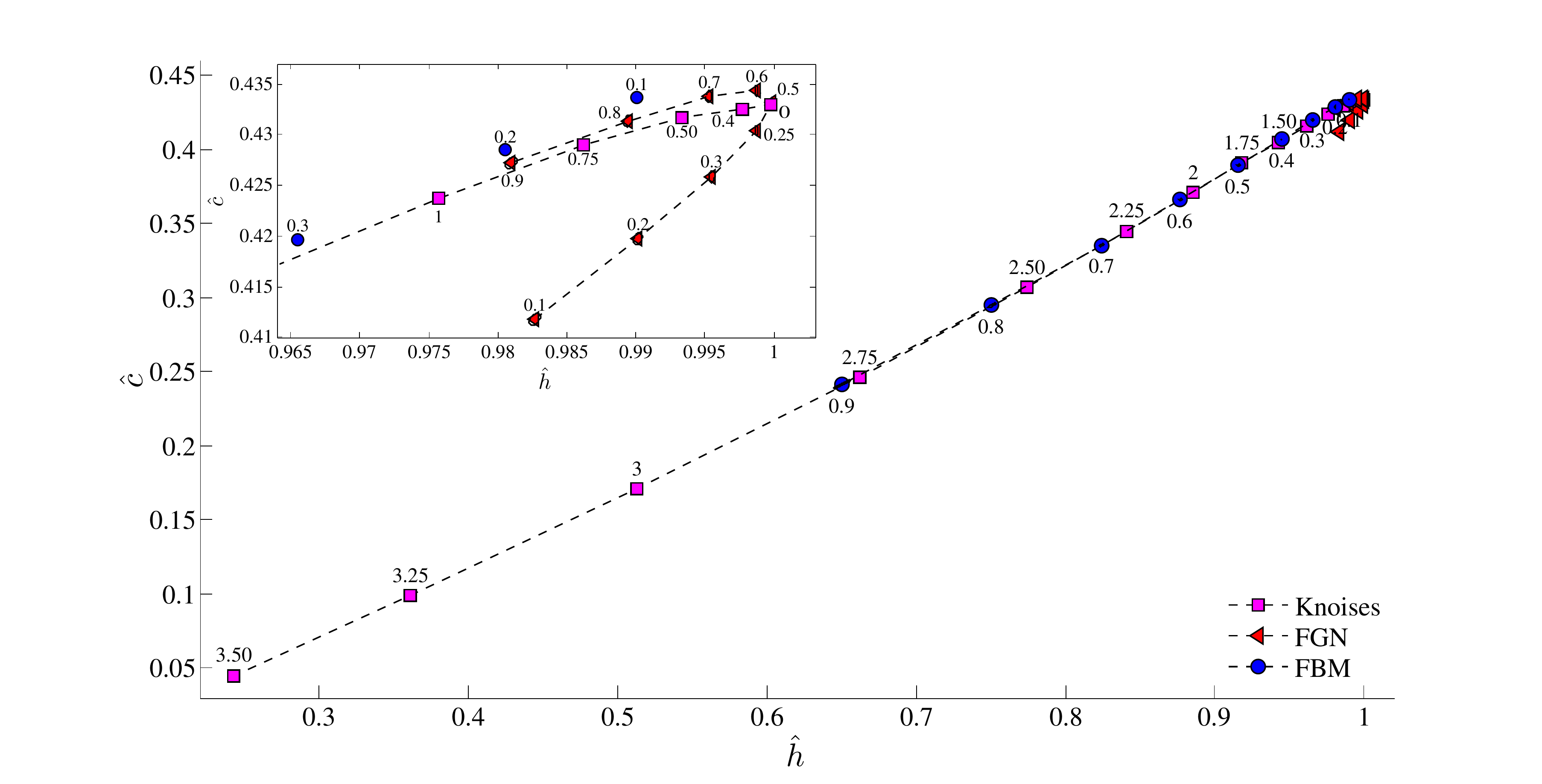}}
\caption{Localization in  the permutation Lempel--Ziv  complexity vs permutation
  Shannon entropy plane of $K$-noise, FGN and FBM sequences. The $K$-noises were
  generated using  the parameter $K = 0.25m,  \, m = 1,\ldots,3.5$.  For the FGN
  and FBM we  use the Hurst exponent $H  = 0.1 n, \, n=1,...,9$ .  For each time
  series we  generated $N  = 4.10^4$ times  series of  $L = 10^6$  samples.  The
  corresponding Bandt-Pompe parameter were $d = 5$ and $\tau = 1$. The same kind
  of figure are obtained for the parameter $d = 4, 6$ and $\tau = 1$.}
\label{fig:mapa_FGN_FBM}
\end{figure}

Let us  now analyze  the permutation Lempel--Ziv  complexity vs  the permutation
Shannon entropy for  the $K$-noises, FBM and FGN time  series for various values
of $k$ or $H$. To this aim, we generated $N = 4.10^4$ series of noise, of length
$L = 10^6$  samples each. For the nonGaussian $K$--noises  we used the parameter
$k \in [0 \, ; \, 3.5]$, for the FBM and FGN series the Hurst exponent $H$ is in
$(0,1)$ and the Bandt--Pompe symbolization parameter used for all the series are
$d = 5$ and $\tau = 1$. Figure~\ref{fig:mapa_FGN_FBM} depicts the mean values of
($\widehat{c},~\widehat{h}$) over the realizations, for the each sequences.

As observed  in Figure~\ref{mapa_CyR:fig}, the complexity-entropy  points of the
sequences of the $K$-noises spread along straight line. As intuitively expected,
signal with low  correlation stay in the high entropy  and complexity values, as
zoomed also in the  insert of figure~\ref{fig:mapa_FGN_FBM}. This effect remains
for FBM sequences, but  is no more valid for the FGN;  such a behavior remain to
be more deeply analyzed.
An interesting feature is that the points corresponding to the FBM processes and
$K$-noise remain in an  intermediate-high region on the complexity-entropy plane
while that  corresponding to the FGN are  concentrated in an high  region of the
plane. In particular, given a  spectrum (power-law exponent), the three types of
noise are  clearly separated,  which is  obviously impossible by  the mean  of a
usual  spectral  analysis. This  lets  suggest  that  more than  the  statistics
(Gaussian vs  nonGaussian) and  more than the  stationarity are captured  by the
couple of measures proposed here.

Note that  for the FGN,  as $\min(H,1-H)$ increases,  the absolute value  of the
correlation  increases,  behavior that  is  conserved  in  terms of  permutation
entropy and  permutation complexity. However,  the permutation entropy  given by
$H$ and  $1-H$ are more or  less identical. In  other words, such an  entropy is
unable to  distinguish whether correlation or  anticorrelation characterizes the
underlying  FBM  process.  This  is  revealed  by  the  permutation  Lempel--Ziv
complexity (see the insert of figure~\ref{fig:mapa_FGN_FBM}).
In other  words, the permutation Lempel--Ziv complexity  capture the short-range
correlation,  or persistency  vs non-persistency.   This effect  strengthens the
importance  of using two  different ``complementary''  measures to  analyze such
random sequences.


\section{Discussion}
\label{Discussion:sec}

In the analysis of time series, the challenge of distinguishing chaotic dynamics
from stochastic dynamics underlying an (apparently) complex time series could be
a critical  and subtle issue. Numerous  tools and methods that  attempt to solve
such as  challenge can  be found in  the literature: from  information theoretic
tools  that  address  such  a  problem  from  a  statistics  point  of  view  to
deterministic  analysis  based on  the  notion of  complexity  in  the sense  of
Kolmogorov.  In   this  paper,  we  proposed   to  analyze  time   series  in  a
complexity--entropy  plane.  The  idea   is  to  combine  both  statistical  and
deterministic/algorithmic point of views.

To use these two measures in their natural discrete-state framework, the signals
must  be previously quantized.  In our  approach, the  quantization is  based on
permutation vectors  following the  scheme proposed by  Bandt and  Pompe dealing
with the so-called permutation entropy.

The Shannon  entropy and the  Lempel--Ziv complexity applied to  the permutation
vectors  result  in  the  so-called  \textit{permutation  Shannon  entropy}  and
\textit{permutation  Lempel--Ziv complexity}  respectively.  We  use  the hence
defined complexity-entropy  plane to analyze  several well known time  series of
the  literature  --chaotic  maps,  $K$-noises, fractional  Brownian  motion  and
fractional Gaussian noise. In  particular, the proposed representation allows to
clearly distinguish chaotic maps from random processes, to classify the chaotic
maps  according to their  ``non inverted'',  ``dissipative''and ``conservative''
characteristics,  or  to  separate  noise  sharing the  same  spectrum  capturing
implicitly both their statistics  and their stationarity/nonstationarity. Such a
plane  appears thus  to  be a  good  alternative or  complement  of the  already
proposed maps.




As future direction of investigation, the maps proposed in the literature should
be compared through automatic classification approaches. One can also imagine to
combine three  measures to capture  more complementary aspects rather  than only
``permutation''  statistical  and algorithmic  aspects,  without  a too  complex
estimation/evaluation  procedure.   Similarly,  rather  than   Shannon  entropy,
generalized entropy  may be able to  capture finer statistical  aspects (\eg the
tails of head of distribution via R\'enyi-Tsallis entropies).



\bibliography{CLZ_H}

\begin{thebibliography}{10}

\bibitem{ZozBla03}
S.~Zozor, O.~Blanc, V.~Jacquemet, N.~Virag, J.-M. Vesin, E.~Pruvot,
  L.~Kappenberger, and C.~Henriquez.
\newblock A numerical scheme for modeling wavefront propagation on a monolayer
  of arbitrary geometry.
\newblock {\em IEEE Transactions on Biomedical Engineering}, 50(4):412--420,
  April 2003.

\bibitem{KanPut04}
N.~Kannathal, S.~K. Puthusserypady, and L.~C. Min.
\newblock Complex dynamics of epileptic {EEG}.
\newblock In {\em Proc. of the 26th Annual International Conference of the
  {IEEE} Engineering in Medicine and Biology Society ({EMBS'04})}, volume~1,
  pages 604--607, San Fransisco, CA, USA, September 1-5 2004. IEEE.

\bibitem{Art94}
W.~B. Arthur.
\newblock Inductive reasoning and bounded rationality.
\newblock {\em The American Economic Review}, 84(2):406--411, May 1994.

\bibitem{Tum84}
N.~B. Tuma.
\newblock {\em Social Dynamics Models and methods}.
\newblock Elsevier, 1984.

\bibitem{ShiFis84}
R.~J. Shiller, S.~Fischer, and B.~M. Friedman.
\newblock Stock prices and social dynamics.
\newblock {\em Brookings papers on economic activity}, 15(2):457--510, 1984.

\bibitem{Raj00}
M.~Rajkovi{\'c}.
\newblock Extracting meaningful information from financial data.
\newblock {\em Physica A}, 287(3-4):383--395, December 2000.

\bibitem{PonPro02}
V.~I. Ponomarenko and M.~D. Prokhorov.
\newblock Extracting information masked by the chaotic signal of a time-delay
  system.
\newblock {\em Physical Review E}, 66(2):026215, August 2002.

\bibitem{BroKin86}
D.~S. Broomhead and G.~P. King.
\newblock Extracting qualitative dynamics from experimental data.
\newblock {\em Physica D}, 20(2-3):217--236, June-July 1986.

\bibitem{QuiArn00}
R.~Quian {Q}uiroga, J.~Arnhold, K.~Lehnertz, and P.~Grassberger.
\newblock Kulback--{L}eibler and renormalized entropies: Applications to
  electroencephalograms of epilepsy patients.
\newblock {\em Physical Review E}, 62(6):8380--8386, December 2000.

\bibitem{RosBla01}
O.~A. Rosso, S.~Blanco, J.~Yordanova, V.~Kolev, A.~Figliola, M.~Sch{\"u}rmann,
  and E.~Ba{\c{s}}ar.
\newblock Wavelet entropy: a new tool for analysis of short duration brain
  electrical signals.
\newblock {\em Journal of neuroscience methods}, 105(1):65--75, January 2001.

\bibitem{Sch00}
T.~Schreiber.
\newblock Measuring information transfer.
\newblock {\em Physical Review Letters}, 85(2):461--464, July 2000.

\bibitem{WolSwi85}
A.~Wolf, J.~B. Swift, H.~L. Swinney, and J.~A. Vastano.
\newblock Determining {L}yapunov exponents from a time series.
\newblock {\em Physica D}, 16(3):285--317, July 1985.

\bibitem{Nag02}
R.~Nagarajan.
\newblock Quantifying physiological data with lempel-ziv complexity-certain
  issues.
\newblock {\em Biomedical Engineering, IEEE Transactions on},
  49(11):1371--1373, 2002.

\bibitem{AboHor06}
M.~Aboy, R.~Hornero, D.~Ab{\'a}solo, and D.~{\'A}lvarez.
\newblock Interpretation of the {L}empel-{Z}iv complexity measure in the
  context of biomedical signal analysis.
\newblock {\em IEEE Transactions on Biomedical Engineering}, 53(11):2282--2288,
  November 2006.

\bibitem{ZozRav05}
S.~Zozor, P.~Ravier, and O.~Buttelli.
\newblock On {L}empel--{Z}iv complexity for multidimensional data analysis.
\newblock {\em Physica A}, 345(1-2):285--302, January 2005.

\bibitem{VigBer03}
C.~Vignat and J.-F. Bercher.
\newblock Analysis of signals in the {F}isher-{S}hannon information plane.
\newblock {\em Physics Letters A}, 312(1-2):27--33, June 2003.

\bibitem{RosLar07}
O.~A. Rosso, H.~A. Larrondo, M.~T. Martin, A.~Plastino, and M.~A. Fuentes.
\newblock Distinguishing noise from chaos.
\newblock {\em Physical Review Letters}, 99(15):154102, October 2007.

\bibitem{RosOli13}
O.~A. Rosso, F.~Olivares, L.~Zunino, L.~De Micco, A.~L.~L. Aquino, A.~Plastino,
  and H.~A. Larrondo.
\newblock Characterization of chaotic maps using the permutation
  {B}andt--{P}ompe probability-distribution.
\newblock {\em The European Physics Journal B}, 86(4):116--128, april 2013.

\bibitem{RosCra09}
O.~A. Rosso, H.~Craig, and P.~Moscato.
\newblock Shakespeare and other {E}nglish renaissance authors as characterized
  by information theory complexity quantifiers.
\newblock {\em Physica A}, 388(6):916--926, March 2009.

\bibitem{ZunSor10}
L.~Zunino, M.~C. Soriano, I.~Fischer, O.~A. Rosso, and C.~R. Mirasso.
\newblock Permutation-information-theory approach to unveil delay dynamics from
  time-series analysis.
\newblock {\em Physical Review E}, 82(4):046212, October 2010.

\bibitem{MonRos14}
F.~Montani and O.~A. Rosso.
\newblock Entropy-complexity characterization of brain development in chickens.
\newblock {\em Entropy}, 16(8):4677--4692, 2014.

\bibitem{LamMar04}
P.~W. Lamberti, M.~T. Martin, A.~Plastino, and O.~A. Rosso.
\newblock Intensive entropic non-triviality measure.
\newblock {\em Physica A}, 334(1-2):119--131, March 2004.

\bibitem{LemZiv76}
A.~Lempel and J.~Ziv.
\newblock On the complexity of finite sequences.
\newblock {\em IEEE Transactions on Information Theory}, 22(1):75--81, January
  1976.

\bibitem{Sha48}
C.~E. Shannon.
\newblock A mathematical theory of communication.
\newblock {\em The Bell System Technical Journal}, 27(4):623--656, October
  1948.

\bibitem{BeiDud97}
J.~Beirlant, E.~J. Dudewicz, L.~Gy{\"o}rfi, and E.~C. {van der Meulen}.
\newblock Nonparametric entropy estimation: An overview.
\newblock {\em International Journal of Mathematical and Statistical Sciences},
  6(1):17--39, June 1997.

\bibitem{LeoPro08}
N.~Leonenko, L.~Pronzato, and V.~Savani.
\newblock A class of {R\'e}nyi information estimators for multidimensional
  densities.
\newblock {\em Annals of Statistics}, 36(5):2153--2182, October 2008.

\bibitem{SchGra96}
T.~Sch{\"u}rmann and P.~Grassberger.
\newblock Entropy estimation of symbol sequences.
\newblock {\em Chaos}, 6(3):414, September 1996.

\bibitem{HerMa02}
A.~O. {Hero III\@}, B.~Ma, O.~J.~J. Michel, and J.~Gorman.
\newblock Application of entropic spanning graphs.
\newblock {\em IEEE Signal Processing Magazine}, 19(5):85--95, September 2002.

\bibitem{FrePom07}
S.~Frenzel and B.~Pompe.
\newblock Partial mutual information for coupling analysis of multivariate time
  series.
\newblock {\em Physical Review Letters}, 99(20):204101, November 2007.

\bibitem{Ros56}
M.~Rosenblatt.
\newblock Remarks on some nonparametric estimates of a density function.
\newblock {\em The Annals of Mathematical Statistics}, 27(3):832--837,
  September 1956.

\bibitem{Par62}
E.~Parzen.
\newblock On estimation of a probability density function and mode.
\newblock {\em The Annals of Mathematical Statistics}, 33(3):1065--1076,
  September 1962.

\bibitem{BanPom02}
C.~Bandt and B.~Pompe.
\newblock Permutation entropy: A natural complexity measure for time series.
\newblock {\em Physical Review Letters}, 88(17):174102, April 2002.

\bibitem{ZozMat14}
S.~Zozor, D.~Mateos, and P.~W. Lamberti.
\newblock Mixing {B}andt--{P}ompe and {L}empel--{Z}iv approaches: another way
  to analyze the complexity of continuous-states sequences.
\newblock {\em The European Physical Journal B}, 87(5):107, May 2014.

\bibitem{Bol64}
L.~Boltzmann {(translated by Stephen G. Brush)}.
\newblock {\em Lectures on Gas Theory}.
\newblock Dover, Leipzig, Germany, 1964.

\bibitem{Gib02}
J.~W. Gibbs.
\newblock {\em Elementary Principle in Statistical Mechanics}.
\newblock University Press - John Wilson and son, Cambridge, USA, 1902.

\bibitem{vNeu27}
J.~{von Neumann}.
\newblock Thermodynamik quantenmechanischer gesamtheiten.
\newblock {\em Nachrichten von der Gesellschaft der Wissenschaften zu
  G{\"o}ttingen}, 1:273--291, 1927.

\bibitem{Nie52:v2}
F.~R.~S. W.~D.~Nieven, M.~A.
\newblock {\em The scientific papers of {J}ames {C}lerk {M}axwell}, volume~2.
\newblock Dover, New-York, 1952.

\bibitem{Jay65}
E.~T. Jaynes.
\newblock Gibbs vs {B}oltzmann entropies.
\newblock {\em American Journal of Physics}, 33(5):391--398, May 1965.

\bibitem{MulMul09}
I.~M{\"u}ller and W.~H. M{\"u}ller.
\newblock {\em Fundamentals of Thermodynamics and Applications. With Historical
  Annotations and Many Citations from {A}vogadro to {Z}ermelo}.
\newblock Springer, Berlin, 2009.

\bibitem{Pla15}
M.~Planck.
\newblock {\em Eight Lectures on Theoretical Physics}.
\newblock Columbia University Press, New-York, 2015.

\bibitem{Khi57}
A.~I. Khinchin.
\newblock {\em Mathematical foundations of information theory}.
\newblock Dover Publications, New-York, 1957.

\bibitem{CovTho06}
T.~M. Cover and J.~A. Thomas.
\newblock {\em Elements of Information Theory}.
\newblock John Wiley \& Sons, Hoboken, New Jersey, 2nd edition, 2006.

\bibitem{Dob58}
R.~L. Dobrushin.
\newblock A simplified method of experimentally evaluating the entropy of a
  stationary sequences.
\newblock {\em Theory of Probability and its Applications}, 3(4):428--430,
  1958.

\bibitem{Vas76}
O.~Vasicek.
\newblock A test for normality based on sample entropy.
\newblock {\em Journal of the Royal Statistical Society B}, 38(1):54--59, 1976.

\bibitem{KozLeo87}
L.~F. Kozachencko and N.~N. Leonenko.
\newblock Sample estimate of the entropy of a random vector.
\newblock {\em Problems in Information transmission}, 23(2):95--101, 1987.

\bibitem{ZivLem77}
J.~Ziv and A.~Lempel.
\newblock A universal algorithm for sequential data compression.
\newblock {\em IEEE Transactions on Information Theory}, 23(3):337--343, May
  1977.

\bibitem{ZivLem78}
J.~Ziv and A.~Lempel.
\newblock Compression of individual sequences via variable-rate coding.
\newblock {\em IEEE Transactions on Information Theory}, 24(5):530--536,
  September 1978.

\bibitem{WynZiv89}
A.~D. Wyner and J.~Ziv.
\newblock Some asymptotic properties of the entropy of a stationary ergodic
  data source with applications to data compression.
\newblock {\em IEEE Transactions on Information Theory}, 35(6):1250--1258,
  november 1989.

\bibitem{Han89}
G.~Hansel.
\newblock Estimation of the entropy by the {L}empel-{Z}iv method.
\newblock {\em Lecture Notes in Computer Science (Electronic Dictionaries and
  Automata in Computational Linguistics)}, 377:51--65, 1989.

\bibitem{KasSch87}
F.~Kaspar and H.~G. Schuster.
\newblock Easily calculable measure for the complexity of spatiotemporal
  patterns.
\newblock {\em Physical Review A}, 36(2):842--848, July 1987.

\bibitem{Tak81}
F.~Takens.
\newblock Detecting strange attractors in turbulence.
\newblock In D.~Rand and L.-S. Young, editors, {\em Dynamical Systems and
  Turbulence}, volume 898 of {\em Lecture Notes in Mathematics}, pages
  366--383. Springer Verlag, Warwick, 1981.

\bibitem{Rob11}
J.~C. Robinson.
\newblock {\em Dimensions, Embeddings, and Attractors}.
\newblock Cambridge University Press, Cambdrige, UK, 2011.

\bibitem{GerGra92}
A.~Gersho and R.~M. Gray.
\newblock {\em Vector quantization and signal compression}.
\newblock Kluwer, Boston, 1992.

\bibitem{AmiKoc06}
J.~M. Amig{\'o}, L.~Kocarev, and J.~Szczepanski.
\newblock Order patterns and chaos.
\newblock {\em Physics Letters A}, 355(1):27--31, June 2006.

\bibitem{Ami10}
J.~M. Amig{\'o}.
\newblock {\em Permutation Complexity in Dynamical Systems}.
\newblock Springer Verlag, Heidelberg, 2010.

\bibitem{AmiZam07}
J.~M. Amig{\'o}, S.~Zambrano, and M.~A.~F. Sanju{\'a}n.
\newblock True and false forbidden patterns in deterministic and random
  dynamics.
\newblock {\em Europhysics Letters}, 79(5):50001, September 2007.

\bibitem{RosCar12}
O.~A. Rosso, L.~C. Carpi, P.~M. Saco, and M.~{G\'omez Ravetti}.
\newblock Causality and the entropy-complexity plane: Robustness and missing
  ordinal patterns.
\newblock {\em Physica A}, 391(1-2):42--45, january 2012.

\bibitem{Spr03}
J.C. Sprott.
\newblock {\em Chaos and time-series analysis}.
\newblock Oxford University Press, Oxford, 2003.

\bibitem{DutHor81}
P.~Dutta and P.~M. Horn.
\newblock Low-frequency fluctuations in solids: {$1/f$} noises.
\newblock {\em Reviews of Modern Physics}, 53(3):497--516, July 1981.

\bibitem{Hos81}
J.~R.~M. Hosking.
\newblock Fractional differencing.
\newblock {\em Biometrika}, 68(1):165--176, April 1981.

\bibitem{WesShl90}
B.~West and M.~Shlesinger.
\newblock The noise in natural phenomena.
\newblock {\em American Scientist}, 78(1):40--45, January-February 1990.

\bibitem{Kol40}
A.~N. Kolmogorov.
\newblock Sienersche spiralen und einige andere interessante kurven im
  hilbertschen raum.
\newblock {\em Doklady Akademii nauk SSSR}, 26(2):115--118, 1940.

\bibitem{Hur51}
H.~Hurst.
\newblock Long-term storage capacity in reservoirs.
\newblock {\em Transactions of the American Society of Civil Engeniering},
  116:770--799, 1951.

\bibitem{ManNes68}
B.~Mandelbrot and J.~W. {Van Ness}.
\newblock Fractional {B}rownian motions, fractional noises and applications.
\newblock {\em SIAM review}, 10(4):422--437, October 1968.

\bibitem{Fla89}
P.~Flandrin.
\newblock On the spectrum of fractional {B}rownian motions.
\newblock {\em IEEE Transactions on Information Theory}, 35(1):197--199,
  January 1989.

\bibitem{ZunPer07}
L.~Zunino, D.~.G P{\'e}rez, M.~T. Mart{\'i}n, A.~Plastino, M.~Garavaglia, and
  O.~A. Rosso.
\newblock Characterization of {G}aussian self-similar stochastic processes
  using wavelet-based informational tools.
\newblock {\em Physical Review E}, 75(2):021115, February 2007.

\bibitem{Ber94}
J.~Beran.
\newblock {\em Statistics for Long-Memory Processes}.
\newblock Chapman \& Hall, New-York, 1994.

\bibitem{SamTaq94}
G.~Samorodnitsky and M.~S. Taqqu.
\newblock {\em Stable Non-{G}aussian Random Processes. Stochastic Models with
  infinite Variance}.
\newblock Chapman \& Hall, New-York, 1994.

\bibitem{Fed88}
J.~Feder.
\newblock {\em Fractals}, volume~9.
\newblock Springer, New-York, 1988.

\bibitem{MolLiu97}
F.~J. Molz, H.~H. Lui, and J.~Szulga.
\newblock Fractional {B}rownian motion and fractional {G}aussian noise in
  subsurface hydrology: A review, presentation of fundamental properties, and
  extensions.
\newblock {\em Water Resources Research}, 33(10):2273--2286, October 1997.

\bibitem{Sam07}
G.~Samorodnitsky.
\newblock Long range dependence.
\newblock {\em Foundations and Trends{$^{\mbox{\textregistered}}$} in
  Stochastic Systems}, 1(3):163--257, 2007.

\bibitem{DavHar87}
R.~B. Davies and D.~S. Harte.
\newblock Tests for {H}urst effect.
\newblock {\em Biometrika}, 74(1):95--101, March 1987.

\bibitem{AbrFel96}
P.~Abry and F.~Sellan.
\newblock The wavelet-based synthesis for fractional {B}rownian motion proposed
  by f. sellan and y. meyer: Remarks and fast implementation.
\newblock {\em Applied and Computational Harmonic Analysis}, 3(4):377--383,
  October 1996.

\bibitem{KinRee98}
W.~Kinzel and G.~Reents.
\newblock {\em Physics by Computers -- Programming Physical Problems Using
  {M}athematica{$^{\mbox{\scriptsize{\textregistered}}}$} and {C}}.
\newblock Springer Verlag, Heidelberg, 1998.

\end{thebibliography}
\bibliographystyle{unsrt}
\end{document}